\documentclass{svjour2}                    
\smartqed  
\usepackage{graphicx}

\usepackage{amsmath}
\usepackage{amssymb}
\usepackage{bm}
\usepackage{cite}
\usepackage{color}

\newcommand{\rmd}{\mathrm{d}}
\newcommand{\rme}{\mathrm{e}}

\begin{document}

\title{
Lie algebraic discussions for time-inhomogeneous linear birth-death processes with immigration
}


\author{
Jun Ohkubo
}


\institute{
Jun Ohkubo \at
Graduate School of Informatics, Kyoto University, \\
Yoshida Hon-mach, Sakyo-ku, Kyoto-shi, Kyoto 606-8501, Japan
\email{ohkubo@i.kyoto-u.ac.jp}
}

\date{$\,$}

\maketitle

\begin{abstract}
Analytical solutions for time-inhomogeneous linear birth-death processes 
with immigration are derived.
While time-inhomogeneous linear birth-death processes
without immigration have been studied by using a generating function approach,
the processes with immigration are here analyzed by Lie algebraic discussions.
As a result,
a restriction for time-inhomogeneity of the birth-death process
is understood 
from the viewpoint of the finiteness of the dimensionality of the Lie algebra.
\end{abstract}

\section{Introduction}

A birth-death process is one of the basic tools to investigate
various stochastic phenomena,
and actually many works have been performed in various research fields
ranging from physics, biology and engineering to social sciences \cite{Gardiner_book}.
Depending on the choice of the transition rates,
we obtain various types of the birth-death processes,
which are available to modeling diverse stochastic phenomena adequately.
While it goes without saying that the birth-death processes have been used
in various contexts,
there is one important problem;
in general, it is difficult to obtain anaytical solutions
for the birth-death processes,
especially, in time-\textit{inhomogeneous} birth-death processes,
in which the transition rates are time-dependent.

As for some of the time-\textit{homogeneous} birth-death processes,
there is a famous spectral formula, so-called Karlin-McGregor spectral formula,
to describe transition probabilities 
\cite{Lederman1954,Karlin1955,Karlin1957a,Karlin1957b,Karlin1958,Schoutens_book}.
In the spectral formula, 
each birth-death process is connected to the three-term relation of a certain type
of an orthogonal polynomial,
and the characteristics of orthogonal polynomials play important roles to obtain the Karlin-McGregor spectral formula.
On the other hand, a generating function approach is another famous analytical treatment
to investigate the birth-death processes \cite{Gardiner_book}.
As for the time-inhomogeneous cases,
the generating function approach has succeeded in deriving analytical solutions
for a special type of birth-death process \cite{Kendall1948};
the special type of birth-death process, i.e., the linear birth-death process,
has been also discussed in the context of economical issues \cite{Aoki_book}.
In the seminal work by Kendall \cite{Kendall1948},
the linear birth-death process with a specific initial condition (a single ``ancestor'')
has been analyzed, and the analytical solutions are given.
However, it has not been obvious which types of the time-inhomogeneous birth-death processes
can be solved analytically.

In the present paper, analytical solutions for
the time-inhomogeneous linear birth-death processes \textit{with immigration}
are derived.
It will be clarified why
the immigration process, which is not considered in the Kendall's work,
makes the analytical treatment difficult.
While there are works by Branson \cite{Branson1991,Branson2000},
in which the immigration effects for the same specific initial condition
as by Kendall have been discussed a little,
here, complete solutions for arbitrary initial conditions are given
by employing a completely different approach,
i.e., the Lie algebraic approach.
The Lie algebraic approach will easily clarify
the fact that
the time-inhomogeneous immigration rate should be proportional
to the time-inhomogeneous birth rate
in order to derive the analytical solutions.

The outline of the present paper is as follows.
In section 2, the model for the birth-death process is explained.
Section 3 gives basic formulations for the Lie algebraic discussions.
Sections 4 and 5 show the application of the Lie algebraic discussions
to the evaluation of transition probabilities and moments, respectively.
Section 6 gives concluding remarks.

\section{Time-inhomogeneous linear birth-death process with immigration}

In the present paper, the following birth-death process is analyzed:
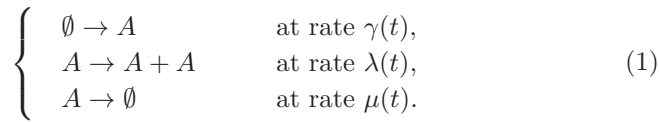
\begin{align}
\begin{cases}
\quad \emptyset \to A \qquad & \textrm{at rate $\gamma(t)$},  \\
\quad A \to A+A \qquad & \textrm{at rate $\lambda(t)$},  \\
\quad A \to \emptyset \qquad & \textrm{at rate $\mu(t)$}.
\end{cases}
\label{eq_process}
\end{align}
The first process in \eqref{eq_process} means the immigration process,
and the seminal work by Kendall \cite{Kendall1948} corresponds
to the case without the immigration, i.e., $\gamma(t) = 0$.
Note that the actual rates of the second and third processes depend
on the number of particles $A$.
In order to see the transition procedures of the model more clearly,
we here give the master equation of the process as follows:
\begin{align}
\frac{\rmd}{\rmd t} P(n,t) 
=&
\gamma(t) P(n-1,t) - \gamma(t) P(n,t) \nonumber \\
&+ \lambda(t) (n-1) P(n-1,t) - \lambda(t) n P(n,t) \nonumber \\
&+ \mu(t) (n+1) P(n+1,t) - \mu(t) n P(n,t),
\label{eq_master_equation}
\end{align}
where $n$ is the number of particle $A$, and $n \in \mathbb{N}$.
The probability $P(n,t)$ means that there is $n$ particles at time $t$.
In the works by Kendall \cite{Kendall1948} 
and Branson \cite{Branson1991,Branson2000},
only the specific initial condition with a single ``ancestor'',
i.e., $P(1,0) = 1$ and $P(i,0) = 0$ for $i \neq 1$,
has been discussed.
In the present paper, 
the initial condition is not restricted
and the initial value can take an arbitrary non-negative integer value.

When we consider the no-immigration cases ($\gamma(t) = 0$),
$\lambda(t)$ and $\mu(t)$ can take \textit{arbitrary} time-dependency,
as discussed by Kendall \cite{Kendall1948}.
However, for $\gamma(t) \neq 0$ cases,
as shown later,
we must restrict the birth rate, $\lambda(t)$,
and the immigration rate, $\gamma(t)$,
as follows:
\begin{align}
\gamma(t) = \beta \lambda(t),
\label{eq_restriction}
\end{align}
where $\beta$ is a time-independent constant.

\section{Application of the Wei-Norman method and resulting restriction}

In this section, the Lie algebraic approach are explained.
Firstly, an algebraic probabilistic method,
the so-called Doi-Peliti formulation \cite{Doi1976,Doi1976a,Peliti1985}, 
is introduced, which is needed for the Lie algebraic discussions.
Secondly, an application of the Wei-Norman method \cite{Wei1963,Wei1964}
to our problem is discussed.
In addition, for readers' convenience, 
we give a brief summary of the Wei-Norman method in the Appendix.

\subsection{Algebraic probabilistic method: Doi-Peliti formulation}

The Doi-Peliti formulation \cite{Doi1976,Doi1976a,Peliti1985},
is a famous tool in nonequilibrium statistical physics;
reaction-diffusion processes and their critical behaviors
have been mainly studied by employing the Doi-Peliti formulation
and field theoretical methods.
(As for the applications of the Doi-Peliti formulation,
see the review paper \cite{Tauber2005} and the textbook \cite{Altland_book},
for example.)
While the Doi-Peliti formulation can be extended a little
as shown in recent works \cite{Ohkubo2012,Ohkubo2013},
we here employ the conventional formulation, as introduced below.

In the Doi-Peliti formulation,
the following creation and annihilation operators are introduced:
\begin{align}
[ a, a^\dagger] \equiv a a^\dagger - a^\dagger a = 1,
\end{align}
where $a^\dagger$ is the creation operator,
and $a$ corresponds to the annihilation operator.
The actions of the operators on a `ket' state $|n\rangle$
are defined as follows:
\begin{align}
a^\dagger |n\rangle = |n+1\rangle, \quad a |n\rangle = n |n-1 \rangle,
\quad a | 0 \rangle = 0,
\end{align}
where $| 0 \rangle$ is the vacuum state.
As the dual state,
the following `bra' state $\langle n |$ is introduced:
\begin{align}
\langle n| a^\dagger = \langle n-1 | n , \quad
\langle n| a = \langle n+1 |, \quad
\langle 0| a^\dagger = 0.
\end{align}
The bra and ket states satisfy the following inner product:
\begin{align}
\langle m | n \rangle = n! \delta_{m,n}.
\end{align}

The key point in the Doi-Peliti formulation
is the rewriting of the original master equation \eqref{eq_master_equation}
in terms of the creation and annihilation operators.
Here, we introduce the following state vector $| \varphi(t) \rangle$:
\begin{align}
| \varphi(t) \rangle = \sum_{n=0}^\infty P(n,t) | n \rangle,
\label{eq_state_vector}
\end{align}
in which $P(n,t)$ is the probability in the original master equation \eqref{eq_master_equation}.
Then, the following time-evolution equation
for the state vector $| \varphi(t)\rangle$
is derived from the original master equation:
\begin{align}
\frac{\mathrm{d}}{\mathrm{d} t} | \varphi(t) \rangle = H(t) | \varphi(t) \rangle,
\label{eq_time_evolution_phi}
\end{align}
where
\begin{align}
H(t) = \gamma(t) (a^\dagger - I) + \lambda(t) (a^\dagger a^\dagger a - a^\dagger a)
+ \mu(t) (a - a^\dagger a).
\label{eq_generator}
\end{align}
(It is easy to confirm that Eq.~\eqref{eq_time_evolution_phi}
corresponds to the original master equation \eqref{eq_master_equation}
by comparing each coefficient of $| n \rangle$.)

Since Eq.~\eqref{eq_time_evolution_phi} is described
only in terms of the creation and annihilation operators,
one can employ conventional field-theoretical method,
such as the coherent-state path integral method (for example, see \cite{Tauber2005}).
In order to use the coherent-state path integral method,
usually some more definitions and preparations are needed.
However, here, the formulations in Eqs.~\eqref{eq_time_evolution_phi}
and \eqref{eq_generator}  are enough
for the following Lie algebraic discussions.

\subsection{Application of the Wei-Norman method}

The Wei-Norman method \cite{Wei1963,Wei1964}
is one of the method to treat linear differential equations with time-varying coefficients.
For example, the Wei-Norman method
has been used to analyze Fokker-Planck equations \cite{Wolf1988}
and financial issues \cite{Lo2001}.
In Ref.~\cite{House2012},
chemical reaction systems has been treated via the Wei-Norman method,
but a kind of infinite-matrix formulation has been used.
The infinite-matrix formulation is a little difficult to treat,
and then the Doi-Peliti formulation would be more suitable
for the Wei-Norman method.
Actually, in a recent work by the author of the present paper,
the Wei-Norman method has been used
to analyze a simple birth-death process in combination with the Doi-Peliti formulation  \cite{Ohkubo2014}.

The aim here is focusing on the Lie algebraic structure and its consequences,
and hence the explanation of the Wei-Norman method is omitted;
see the Appendix for the brief summary of the Wei-Norman method.

The most important condition for the Wei-Norman method
is the fact that the Lie algebra has a \textit{finite} dimension.
For example, in the no-immigration cases ($\gamma(t) = 0$),
the generator $H(t)$ in Eq.~\eqref{eq_generator}
consists of the following operators:
\begin{align*}
a^\dagger a^\dagger a, \quad a^\dagger a, \quad  a.
\end{align*}
Since the commutation relations for the above three operators are closed,
i.e.,
\begin{align*}
[a^\dagger a^\dagger a, a] = -2 a^\dagger a, \quad 
[a^\dagger a^\dagger a, a^\dagger a] = - a^\dagger a^\dagger a, \quad
[a, a^\dagger a] = -a,
\end{align*}
the Wei-Norman method is applicable.
(Note that the Lie algebra composed with the three operators,
$\{a^\dagger a^\dagger a, a, a^\dagger a\}$
is not \textit{solvable},
and hence it could not be guaranteed that
the analytical solution is obtained by quadrature.
As for such discussions based on the Lie algebraic structures,
see the original papers by Wei and Norman \cite{Wei1963,Wei1964}.
However, as shown later,
we will see that analytical solutions for the birth-death process 
can be obtained in our restricted cases.)
Next, let us tackle with the general case with immigration.
In this case, we have the following five operators:
\begin{align*}
a^\dagger, \quad I, \quad a^\dagger a^\dagger a, \quad a^\dagger a, \quad a.
\end{align*}
The problem is as follows:
Is it possible to consider the following Lie algebra
in order to employ the Wei-Norman method?:
\begin{align*}
\mathcal{L} \overset{\textrm{\large ?}}{=}  \left\{
I, \,\, a^\dagger, \,\, a, \,\, a^\dagger a, \,\, a^\dagger a^\dagger a
\right\}.
\end{align*}
Here, note that the Lie algebra does not have a finite dimension
because 
\begin{align}
[a^\dagger a^\dagger a, a^\dagger] = a^\dagger a^\dagger,
\end{align}
and hence a new operator $a^\dagger a^\dagger$ must be included in the Lie algebra,
and $[a^\dagger a^\dagger a, (a^\dagger)^2] = 2 (a^\dagger)^3$, and so on.
The important point here is that from the above naive construction,
we cannot have a Lie algebra with a finite dimension.

Here, the following trick is employed \cite{Miki2012}:
using a time-independent constant $\beta$,
consider the following Lie algebra:
\begin{align}
\mathcal{L} = \left\{
I, \,\, a, \,\, a^\dagger a, \,\, a^\dagger (\beta + a^\dagger a)
\right\}.
\label{eq_closed_Lie_algebra}
\end{align}
Since
\begin{align}
[a^\dagger (\beta + a^\dagger a), a] = - \beta I - 2 a^\dagger a, \qquad
[a^\dagger (\beta + a^\dagger a), a^\dagger a] = - a^\dagger (\beta + a^\dagger ),
\end{align}
the Lie algebra $\mathcal{L}$ in Eq.~\eqref{eq_closed_Lie_algebra}
is closed,
and hence it becomes possible to apply the Wei-Norman method.

According to the Lie algebra $\mathcal{L}$,
the generator $H(t)$ in Eq.~\eqref{eq_generator}
should be rewritten as follows:
\begin{align}
H(t) &= - \gamma(t) I + \lambda(t) a^\dagger (\beta + a^\dagger a) 
+ \mu(t) a - (\lambda(t) - \mu(t)) a^\dagger a \nonumber \\
&= \sum_{k=1}^4 a_k(t) H_k
\label{eq_generator_new},
\end{align}
where
\begin{align*}
&a_1(t) = - \gamma(t), \, a_2(t) = \lambda(t), \, a_3(t) = \mu(t), \, a_4(t) = - (\lambda(t) - \mu(t)), \\
&H_1 = I, \,\, H_2 = a^\dagger (\beta + a^\dagger a ), \,\, H_3 = a, \,\, H_4 = a^\dagger a,
\end{align*}
and $\beta$ is defined as Eq.~\eqref{eq_restriction},
i.e., $\beta = \gamma(t) / \lambda(t)$

From the above discussions, it is clarified that
for the time-inhomogeneous linear birth-death process with immigration
can be analytically treated (at least, via the Wei-Norman method)
only if the immigration rate $\gamma(t)$ is proportional to the birth rate $\lambda(t)$.
Actually, this is consistent with the previous work by Branson \cite{Branson1991,Branson2000}.
While in the previous work \cite{Branson1991,Branson2000}
the reason why this limitation should be employed has not been stated explicitly,
we here explicitly see the reason via the Lie algebraic discussions.
Of course, as for the death rate $\mu(t)$, an arbitrary time-inhomogeneous function is available.

The remaining task is to apply the Wei-Norman method
to our problem, which needs some tedious calculations,
and the results are as follows:
The following form of the time-evolution operator $U(t)$,
which satisfies $\frac{\rmd}{\rmd t} U(t) = H(t) U(t)$ and $U(0) = I$, is assumed:
\begin{align}
U(t) = \rme^{g_1(t) I} \rme^{g_2(t) a^\dagger (\beta + a^\dagger a)} \rme^{g_3(t) a} \rme^{g_4(t)a^\dagger a},
\label{eq_time_evolution_operator}
\end{align}
and then via the Wei-Norman method,
the following simultaneous ordinary differential equations are derived for each operator in 
the Lie algebra $\mathcal{L}$:
\begin{align}
&(H_1 = I): \nonumber \\
&\qquad -\gamma(t) = \dot{g}_1(t) - \beta g_2(t) \dot{g}_3(t) - \beta g_2(t) g_3(t) \dot{g}_4(t), \\
&(H_2 = a^\dagger (\beta + a^\dagger a) ) : \nonumber \\
&\qquad \lambda(t) = \dot{g}_2(t) + (g_2(t))^2 \dot{g}_3(t) - g_2(t) \dot{g}_4(t) + (g_2(t))^2 g_3(t) \dot{g}_4(t),\\
&(H_3 = a) : \nonumber \\
&\qquad \mu(t) = \dot{g}_3(t) + g_3(t) \dot{g}_4(t),\\
&(H_4 = a^\dagger a) : \nonumber \\
&\qquad - \lambda(t) - \mu(t) = \dot{g}_4(t) - 2 g_2(t) \dot{g}_3(t) - 2 g_2(t) g_3(t) \dot{g}_4(t),
\end{align}
that is,
\begin{align}
&\dot{g}_1(t) = \beta ( \mu(t) g_2(t)  - \lambda(t) ), \\
&\dot{g}_2(t) = \lambda(t) + g_2(t) ( g_2(t) \mu(t) - \mu(t) - \lambda(t)), \label{eq_Riccati}\\
&\dot{g}_3(t) = \mu(t) + g_3(t) (\lambda(t) + \mu(t) - 2 \mu(t) g_2(t)), \\
&\dot{g}_4(t) = - \lambda(t) - \mu(t) + 2 \mu(t) g_2(t).
\end{align}
Equation~\eqref{eq_Riccati} has the form of the Riccati equation,
and using the initial conditions $g_1(0) = g_2(0) = g_3(0) = g_4(0)$,
we finally obtain the following analytical solutions for $\{g_i(t)\}$:
\begin{align}
&g_1(t) = \beta \left(
\rho(t) - \int_0^t  \frac{\mu(\tau)}{W(\tau)} \rmd \tau
\right), \\
&g_2(t) = 1 - \frac{1}{W(t)}, \\
&g_3(t) = \rme^{\rho(t)} (W(t))^2 \int_0^t \frac{\rme^{-\rho(\tau)}}{(W(\tau))^2} \mu(\tau) \rmd \tau, \\
&g_4(t) = - \rho(t) - 2 \ln W(t),
\end{align}
where
\begin{align}
&\rho(t) = \int_0^t (\mu(\tau) - \lambda(\tau)) \rmd \tau, \\
& W(t) = \rme^{- \rho(t)} \left(
1 + \int_0^t \rme^{\rho(\tau)} \mu(\tau) \rmd \tau
\right).
\end{align}

In the following sections,
as examples of some statistical quantities related to the birth-death process,
the transition probabilities from arbitrary initial conditions
and first-order moment with a specific initial condition are derived.

\section{Evaluation of transition probabilities}

Using the Doi-Peliti formulation,
the transition probability from state $n$ at time $t = 0$
to state $m$ at time $t$ is written as 
\begin{align}
P_{n \to m}(t) &= \frac{1}{m!} \langle m | U(t) | n \rangle,
\end{align}
where $U(t)$ is the time-evolution operator in Eq.~\eqref{eq_time_evolution_operator}.
Hence, we have
\begin{align}
P_{n \to m}(t)
&= \frac{1}{m!} \langle m |
\rme^{g_1(t) I} \rme^{g_2(t) a^\dagger (\beta + a^\dagger a)} \rme^{g_3(t) a} \rme^{g_4(t)a^\dagger a}
| n \rangle \nonumber \\
&=  \frac{1}{m!} \rme^{g_1(t) + g_4(t) n} \langle m |
 \rme^{g_2(t) a^\dagger (\beta + a^\dagger a)} \rme^{g_3(t) a} 
| n \rangle.
\end{align}
Here, we focus on the following relations:
\begin{align}
&\langle m | \rme^{g_2(t) a^\dagger (\beta + a^\dagger a)} \nonumber \\
&= \langle m | \sum_{i=0}^\infty \frac{(g_2(t))^i}{i!} \{ a^\dagger (\beta + a^\dagger a) \}^i \nonumber \\
&= \sum_{i=0}^m \frac{(g_2(t))^i}{i!}  \langle m - i | 
\{ m(\beta+m-1) \}  \cdots \{ (m-i+1) (\beta +m-i) \} \nonumber \\
&= \sum_{i=0}^m \frac{(g_2(t))^i}{i!}  \langle m - i | 
\frac{ m! \Gamma(\beta + m)}{(m-i)! \Gamma(\beta + m - i)},
\end{align}
and hence the following transition probabilities are obtained:
\begin{align}
P_{n \to m}(t) \nonumber \\
&\hspace{-15mm} = \begin{cases}
\displaystyle
\sum_{i=0}^m \frac{n! \Gamma(\beta+m)}{i!(m-i)! (n-m+i)! \Gamma(\beta+m-i)}
(g_2(t))^i (g_3(t))^{n-m+i} \qquad (n \ge m),\\
\displaystyle
\sum_{i=0}^n \frac{n! \Gamma(\beta+m)}{i!(n-i)! (m-n+i)! \Gamma(\beta+n-i)}
(g_2(t))^{m-n+i} (g_3(t))^{i} \qquad (n < m).
\end{cases}
\end{align}

\section{Evaluation of moments}

In order to evaluate the moments for the number of particles in the birth-death process at time $t$,
the following projection state is useful (see, for example, \cite{Tauber2005}):
\begin{align}
\langle \mathcal{P} | \equiv \langle 0 | \rme^{a}.
\end{align}
Using the projection state,
the $j$-th order moment is evaluated as follows:
\begin{align}
\overline{n^j} (t) = \sum_{n=0}^\infty n^j P(n,t)
= \langle \mathcal{P} | (a^\dagger a)^j | \varphi(t) \rangle.
\end{align}
While it would be possible to derive higher-order moments with arbitrary initial conditions,
it needs long and tedious calculations;
we here demonstrate the calculation of the first order moment
with the specific initial condition with $n = 1$ at time $t = 0$,
which corresponds to the previous work by Kendall \cite{Kendall1948} when $\gamma(t) = 0$.

Firstly, we employ the following identity:
\begin{align}
\langle 0 | \rme^{a} a^\dagger 
&= \langle 0 | \sum_{i=0}^\infty \frac{1}{i!} a^i a^\dagger \nonumber \\
&= \langle 0 | \sum_{i=0}^\infty \frac{1}{i!} 
\left( i a^{i-1} + a^\dagger a^i \right) \nonumber \\
&=  \langle 0 | \rme^{a}.
\end{align}
Note that $\langle 0 | a^\dagger = 0$.
Hence, we have
\begin{align}
&\overline{n}(t) \nonumber \\
&= \langle \mathcal{P} | (a^\dagger a ) U(t) | 1 \rangle \nonumber \\
&= \langle \mathcal{P} | (a^\dagger a ) 
\rme^{g_1(t) I} \rme^{g_2(t) a^\dagger (\beta + a^\dagger a)} \rme^{g_3(t) a} \rme^{g_4(t)a^\dagger a} 
| 1 \rangle \nonumber \\
&= \rme^{g_1(t) + g_4(t)}
\langle \mathcal{P} |  a \rme^{g_2(t) a^\dagger (\beta + a^\dagger a)}
|1\rangle \nonumber \\
&\quad + g_3(t) \rme^{g_1(t) + g_4(t)} 
\langle \mathcal{P} |  a \rme^{g_2(t) a^\dagger (\beta + a^\dagger a)} | 0 \rangle,
\label{eq_moment_tmp}
\end{align}
where $|1\rangle = a^\dagger | 0 \rangle$ corresponds to the specific initial condition,
i.e., $| \varphi(0) \rangle = P(1,0) | 1 \rangle = | 1 \rangle$.

Secondly, we employ the following two identities:
\begin{align}
\langle \mathcal{P} | a^\dagger = \langle \mathcal{P} |
\end{align}
and
\begin{align}
(a^\dagger a^\dagger a)^{i} a^\dagger |0 \rangle 
= i! (a^\dagger)^{i+1} | 0 \rangle.
\end{align}
Hence, the first term in the last line in Eq.~\eqref{eq_moment_tmp} becomes
(except for the factor $\exp(g_1(t) + g_4(t))$)
\begin{align}
\langle \mathcal{P}  | a \rme^{g_2(t) a^\dagger (\beta + a^\dagger a)} | 1 \rangle 
&= \langle \mathcal{P} |  a \sum_{i=0}^\infty (g_2(t))^i (a^\dagger)^{i+1} | 0 \rangle \nonumber \\
&= \langle \mathcal{P}  |  \sum_{i=0}^\infty (g_2(t))^i (i+1)  (a^\dagger)^i| 0 \rangle \nonumber \\
&= \sum_{i=0}^\infty (g_2(t))^i (i+1)  \nonumber \\
&= \frac{g_2(t)}{(1-g_2(t))^2} + \frac{1}{1-g_2(t)}.
\end{align}

Thirdly, employing the following identity
\begin{align}
(a^\dagger (\beta +  a^\dagger a))^{i} |0 \rangle 
= \beta(\beta+1)\cdots (\beta+i-1) (a^\dagger)^{i} | 0 \rangle,
\end{align}
the second term in the last line in Eq.~\eqref{eq_moment_tmp} becomes
(except for the factor $g_3(t) \exp(g_1(t) + g_4(t))$)
\begin{align}
\langle \mathcal{P}  | a \rme^{g_2(t) a^\dagger (\beta + a^\dagger a)} | 0 \rangle 
&= \langle \mathcal{P} |  a \sum_{i=0}^\infty \frac{(g_2(t))^i }{i!}
\frac{\Gamma(\beta+i)}{\Gamma(\beta)} (a^\dagger)^{i} | 0 \rangle \nonumber \\
&= \langle \mathcal{P} |  \sum_{i=0}^\infty \frac{(g_2(t))^i }{i!}
\frac{\Gamma(\beta+i)}{\Gamma(\beta)} i (a^\dagger)^{i-1} | 0 \rangle \nonumber \\
&=   \sum_{i=0}^\infty \frac{(g_2(t))^i }{(i-1)!}
\frac{\Gamma(\beta+i)}{\Gamma(\beta)}  \nonumber \\
&= \frac{\beta}{(1-g_2(t))^{\beta+1}}.
\end{align}
Finally, we obtain
\begin{align}
\overline{n}(t) = \rme^{g_1(t) + g_4(t)}
(1-g_2(t))^{-\beta-2} \left(
1 + \beta g_2(t) + \beta g_2(t) (1-g_2(t)) g_3(t)
\right).
\end{align}
When $\beta = 0$, the first moment is written as
\begin{align}
\overline{n}(t) = \rme^{- \rho(t)},
\end{align}
which recovers the result by Kendall \cite{Kendall1948}.

\section{Concluding remarks}

The present work corresponds to a little extension of the seminal work by Kendall \cite{Kendall1948}.
Employing a completely different approach from the Kendall's work,
we can adequately recover the previous results.
Moreover, the general results with the immigration effects are given.
While it has not been clarified whether
analytical solutions can be obtained for general time-inhomogeneous cases
or not,
the Lie algebraic discussions
suggest naturally the restriction,
i.e., the time-inhomogeneous immigration rate should be proportional to 
the time-inhomogeneous birth rate.
Note that the discussion is not a proof for the impossibility to obtain analytical solutions
for arbitrary time-inhomogeneity,
but it seems natural that the infinite dimensionality of the Lie algebra 
is related to the impossibility.

The discussions based on the Lie algebraic method
could state the limitation for obtaining analytical solutions
for more general birth-death processes.
For birth-death processes with infinite state space, i.e., the particle number $n$
can take an arbitrary non-negative integer value,
it would be difficult to obtain the Lie algebra with a finite dimension in general.
Although there are some simple systems with multiple species,
which give the Lie algebra with a finite dimension,
the linear birth rate, which corresponds to the process $A \to A + A$,
makes in general the Lie algebraic treatment difficult, as shown in the present paper.
In this sense, the birth-death process without immigration
could be the utmost time-inhomogeneous birth-death process with arbitrary time-dependent rates;
even in the birth-death process with immigration,
there is the restriction for the time-dependent rates.

\begin{acknowledgements}
This work was supported in part by grant-in-aid for scientific research (Grants No.~25870339)
from the Ministry of Education, Culture, Sports, Science and Technology, Japan.
\end{acknowledgements}

\appendix

\section{The Wei-Norman method}

For readers' convenience, the Wei-Norman method is briefly explained in this appendix.
The Wei-Norman method is one of the Lie algebraic method
to solve linear differential equations with varying coefficients.
As for the related algebraic method, the so-called Magnus expansion,
see the review paper in \cite{Blanes2009}.
For the details of the Wei-Norman method, see the original papers \cite{Wei1963,Wei1964}.

Let $\mathcal{L}$ be a finite-dimensional Lie algebra generated by $H_1, \dots, H_L$ under the commutator product.
Note that the following procedures are applicable only if the Lie algebra has a \textit{finite} dimension.

For later use, we define an adjoint operator, $\mathrm{ad}$,
which is a linear operator on $\mathcal{L}$ and
\begin{align}
&(\mathrm{ad} H_i) H_j \equiv [H_i, H_j] = H_i H_j - H_j H_i,\\
&(\mathrm{ad} H_i)^2 H_j = [H_i, [H_i, H_j] ],
\end{align}
and so on.

Define a time-evolution operator $U(t)$,
which satisfies
\begin{align}
\frac{\rmd }{\rmd t} U(t)  = H(t) U(t)
\label{eq_app_time_evolution_equation_for_U_1}
\end{align}
and $U(0) = I$, where $I$ is the identity operator.
In addition, the operator $H(t)$ is assumed to be written as
\begin{align}
H(t) = \sum_{k=1}^K a_k(t) H_k,
\label{eq_app_H}
\end{align}
where $K$ is finite and $K \le L$.

The Wei-Norman method finds an expression of the time-evolution operator $U(t)$
with the following form:
\begin{align}
U(t) = \exp\left( g_1(t) H_1 \right) \exp\left( g_2(t) H_2 \right) \cdots \exp\left( g_L(t) H_L \right),
\label{eq_app_form_U}
\end{align}
where $g_l(0) = 0$ for all $l \in \{1,2,\dots, L\}$.
The time derivative of Eq.~\eqref{eq_app_form_U} gives
\begin{align}
\frac{\rmd }{\rmd t} U(t)  = \sum_{l=0}^L \dot{g}_l(t) 
\left( \prod_{j=1}^{l-1} \exp(g_j(t) H_j) \right) H_i \left( \prod_{j=i}^{L} \exp(g_j(t) H_j) \right).
\label{eq_app_time_evolution_equation_for_U_2}
\end{align}
Performing a post-multiplication by the inverse operator $U^{-1}$, 
and employing the Baker-Campbell-Hausdorff formula,
\begin{align}
\rme^{H_i} H_j \rme^{- H_i} = \rme^{(\mathrm{ad} H_i)} H_j,
\end{align}
the following expression is obtained:
\begin{align}
\left( \frac{\rmd }{\rmd t} U(t) \right) U^{-1}(t)
= \sum_{l=0}^L \dot{g}_l(t) 
\left( \prod_{j=1}^{l-1} \exp\left( g_j(t) (\mathrm{ad} H_j) \right) \right) H_l.
\label{eq_app_time_evolution_equation_for_U_3}
\end{align}
On the other hand, from Eqs.~\eqref{eq_app_time_evolution_equation_for_U_1} and \eqref{eq_app_H},
we have
\begin{align}
\frac{\rmd }{\rmd t} U(t)  = \sum_{l=0}^L a_l(t) H_l U(t),
\label{eq_app_time_evolution_equation_for_U_4}
\end{align}
where $a_l(t) \equiv 0$ for $l > K$.
Hence, the following expression is obtained:
\begin{align}
\left( \frac{\rmd }{\rmd t} U(t) \right)U^{-1}(t)  = \sum_{l=0}^L a_l(t) H_l.
\label{eq_app_time_evolution_equation_for_U_5}
\end{align}
Comparing Eqs.~\eqref{eq_app_time_evolution_equation_for_U_3} with \eqref{eq_app_time_evolution_equation_for_U_5},
we finally obtain
\begin{align}
\sum_{l = 0}^L a_l(t) H_l = \sum_{l=0}^L \dot{g}_l(t) 
\left( \prod_{j=1}^{l-1} \exp\left( g_j(t) (\mathrm{ad} H_j) \right) \right) H_l.
\label{eq_app_result_wei_norman}
\end{align}
That is, we have a linear relation between $a_l(t)$ and $\dot{g}_l(t)$.
Hence, comparing the coefficients of each $H_l$ in the left and right hand sides,
the coupled ordinary differential equations
for $\{\dot{g}_l(t)\}$ are derived.



\end{document}